# Direct optical interfacing of CVD diamond for deported NV-based sensing experiments

**Ludovic Mayer[1], Thierry Debuisschert[*,1]**

[1] Thales Research & Technology, 1 avenue Augustin Fresnel, 91767 Palaiseau cedex, France

Nitrogen-vacancy (NV) defect in diamond is a very promising tool for numerous sensing applications like magnetometry or thermometry. In this paper, we demonstrate a compact and convenient device for magnetic field imaging where a commercial single mode photonic crystal fibre is directly coupled to a commercial CVD ppm diamond. We managed to excite and detect efficiently the luminescence from an ensemble of NV centres and also to perform Electron Spin Resonance (ESR) experiments where the NV hyperfine structure is perfectly resolved under continuous excitation and measurement.

**1 Introduction** In recent years, negatively charged nitrogen vacancy (NV⁻) colour centres in diamond have shown themselves promising candidates as highly sensitive, atomic-sized probes for magnetic field [1], electric field [2], pressure [3] or local temperature [4] measurements. A key point regarding the building of a practical device lies on the integration of the NV-sensor itself in the most compact and flexible way. One solution consists in the use of one single optical fibre to perform both optical excitation and collection of the luminescence signal coming from the NV centres, which allows electron spin to be manipulated and read out. By combining a diamond microcrystal coupled to a 200 µm core diameter optical fibre, Fedotov et al. [5-7] recently built a compact magnetometer based on a similar structure. To improve the sensitivity of this kind of device, bulk CVD diamonds could be used instead of diamond microcrystals since their synthesis is now well mastered but also because intrinsic NV centres exhibit generally far better spin coherence properties. Furthermore, such a system would be easy to combine with specifically engineered CVD diamonds like CVD-growth diamonds along [111] or [113] direction where NV centres exhibit preferential orientation [8, 9] or isotopically purified diamonds with enhanced coherence properties [10]. Another advantage of diamond direct optical interfacing is a more convenient integration especially for NV⁻ magnetometry experiments in cryogenic medium [11-12].

In this work, we present a magnetometer combining a commercial single-mode photonic crystal fibre coupled to a commercial CVD ppm diamond. Taking advantage of the wide spectral acceptance of the fibre, we managed to excite and detect efficiently the luminescence from an ensemble of NV centres and to perform Electron Spin Resonance (ESR) experiments with an ESR linewidth around 550 kHz under continuous excitation and measurement.

**2 Experiment** Optical characterization of the coupling between the diamond and the fibre was performed using a homemade confocal microscopy setup (Fig. 1).

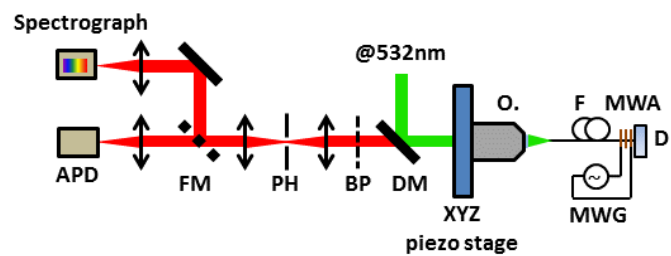

**Figure 1** Sketch of the experimental setup using an inverted confocal microscope to inject and collect light in and from a fibre. D: commercial optical grade CVD diamond (E6); O: air microscope objective (Nachet, 10X, NA = 0.25); DM: dichroic beamsplitter; BP: FF01-697/75 nm band-pass filter (Semrock); PH: 50 µm diameter pinhole; FM: flipping mirror directing the collected photoluminescence either to an imaging spectrograph equipped with a back-illuminated cooled CCD array, or to a silicon avalanche photodiode (APD) in photon-counting regime; F: commercial single-mode photonic bandgap optical fibre (NKT Photonics); MWG: microwave signal generator (SMB 100A, Rhode&Schwarz); MWA: microwave antenna made of 100 µm diameter cooper wire directly coiled around the fibre.

* Corresponding author: e-mail Thierry.debuisschert@thalesgroup.com

For this study, the diamond in which to couple light in and from the optical fibre was a commercial optical grade CVD diamond from E6 containing about 1 ppm of nitrogen impurities. This kind of diamond contains about few ppb of native NV centres which is low enough to avoid dipolar magnetic coupling between neighbour NV centres. In this case, the coherence properties are mainly limited by the interaction between NV centres and the $^{13}$C bath (I-1/2) [13-15].

### 3. Results and discussion

**3.1 Fibre characterization** One of the key points regarding the possibility to optically interface a bulk diamond by the mean of a fibre concerns the fibre characteristics themselves. The fibre has to be able to carry the pump beam at the wavelength 532 nm without leading to excessive parasitic luminescence coming from the core of the fibre. The fibre must also have a large spectral acceptance since the NV$^-$ luminescence spectrum covers the range from 600 nm to 800 nm (Fig. 2).

In our study, we chose a commercial single-mode photonic bandgap fibre from NKT Photonics. This fibre is mainly low-loss from 400 nm to 1700 nm and also possesses a core made of pure silica which limits the parasitic background compared to standard Ge-doped fibre.

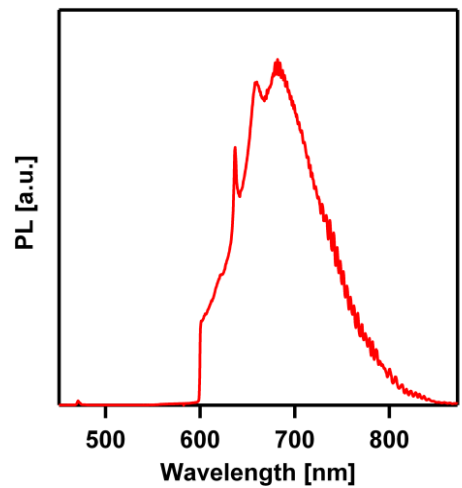

**Figure 2** Luminescence spectrum from HPHT diamond containing 20 ppm of NV centre recording under @ 532 nm illumination. The zero-phonon line at 637 nm and the phonon sidebands are the NV$^-$ signature.

We have first investigated the fibre alone by the mean of a confocal microscope to scan the fibre input (Fig. 3a). Photoluminescence signal was recorded while scanning the fibre input with a 2 mW excitation beam @ 532 nm (Fig. 3b).

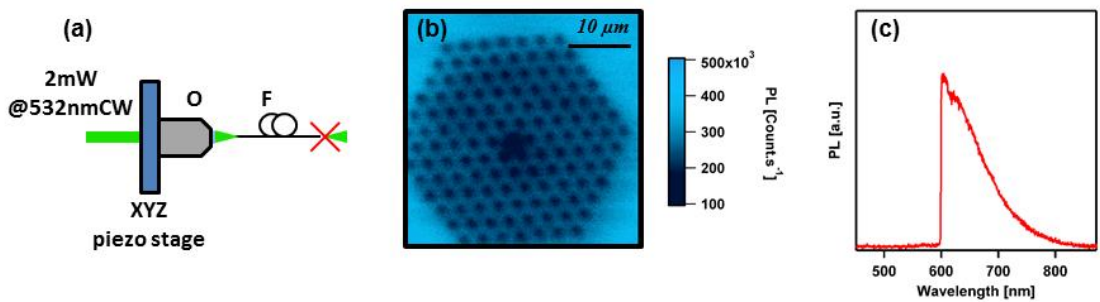

**Figure 3** Optical characterization of the fibre alone. (a) Sketch of the experiment (b) Photoluminescence signal recorded under 2 mW illumination @532nm, 200 pts per line and 3 ms of exposition time per pixel (c) Spectrum of the photoluminescence recorded with the fibre core lying in the focal point of the objective (i.e. maximum beam coupling inside the fibre). The cut-off at wavelength shorter than 600 nm is due to the optical filter.

The photonic crystal structure is clearly resolved and a drop of luminescence is also observed in the middle of the structure corresponding to the case where the pump beam is efficiently coupled into the fibre. Spectral analysis of this low backscattered light in this coupling condition (Fig. 3c) shows a broad spectrum slightly shifted to the blue comparing to the NV$^-$ luminescence spectrum (Fig. 2).

The parasitic luminescence of the fibre related to the pump beam coupling remains low and spectrally shifted compared to NV- spectrum.

**3.2 Study of the fibre/diamond coupling** In order to investigate the coupling between the fibre and a diamond crystal, we performed an experiment similar to the one performed to characterize the fibre alone, but here we put the output of the fibre in direct contact with a CVD diamond crystal (Fig. 4a).

In this configuration, scanning the input port of the fibre leads to the same kind of image than the one obtained previously for the fibre alone but this time a very bright spot of luminescence comes out from the fibre when the pump beam is coupled into the fibre core (Fig. 4b). Spectral analysis of this luminescence signal is given on Fig. 4c. The peak corresponding to the 637 nm zero-phonon line and phonon sidebands of NV- can be clearly observed. The other peak @532nm corresponds to the partially filtered reflection of the pump beam at the fibre-diamond interface. NV centres contained into the diamond can be efficiently excited and their luminescence collected by the mean of this photonic bandgap fibre. In our experimental conditions, the signal (1.5 Mct.s$^{-1}$, 2 mW@532nm) to noise (130 kct.s$^{-1}$, 2 mW@532nm) ratio is about 10 which is very similar to the one obtained for direct microscope imaging of a CVD diamond. Furthermore one can note that, although here the NV concentration and the pump intensity are relatively low, the high count rate observed suggests that the coupling between proximal NV centres and the fibre is efficient.

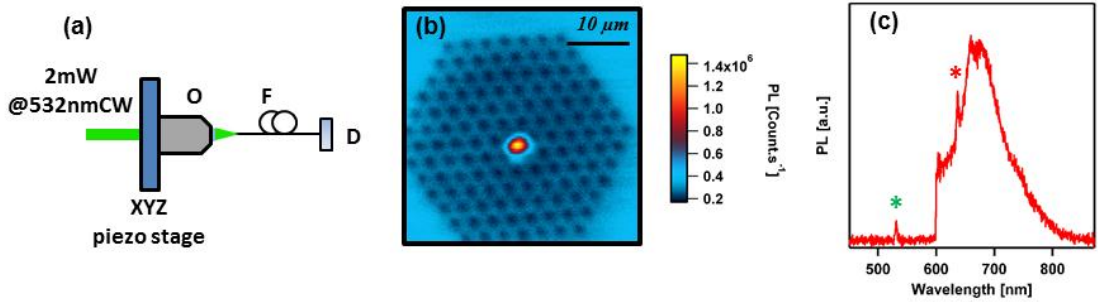

**Figure 4** Optical characterization of the CVD diamond interfaced with a fiber. (a) Sketch of the experiment (b) Photoluminescence signal recorded under 2 mW illumination @532 nm, 200 pts per line and 3 ms of exposition time per pixel (c) Spectrum of the photoluminescence recorded with the fibre core lying in the focal point of the objective (i.e. maximum beam coupling inside the fibre). NV- ZPL @637 nm (red star) and residual pump @532 nm (green star) are highlighted.

**3.3 ESR measurements through a fibre** The NV- defect ground state is a spin triplet with a zero-field splitting $D = 2.87$ GHz between a singlet state $m_s = 0$ and a doublet $m_s = \pm 1$, where $m_s$ denotes the spin projection along the intrinsic quantization axis of the NV defect (Fig. 5a). Note that strain-induced splitting of the $m_s = \pm 1$ spin sublevels has been omitted for clarity purpose and because its contribution is very low especially considering the case of CVD diamond which is the type of diamond used in this study. Under optical illumination, the NV- defect is efficiently polarized into the $m_s = 0$ spin sublevel due to spin selective intersystem crossing. Furthermore, since intersystem crossings are non-radiative, the NV- defect photoluminescence is significantly higher when the $m_s = 0$ state is populated [16]. These combined properties enable the detection of electron spin resonance (ESR) by purely optical means [17].

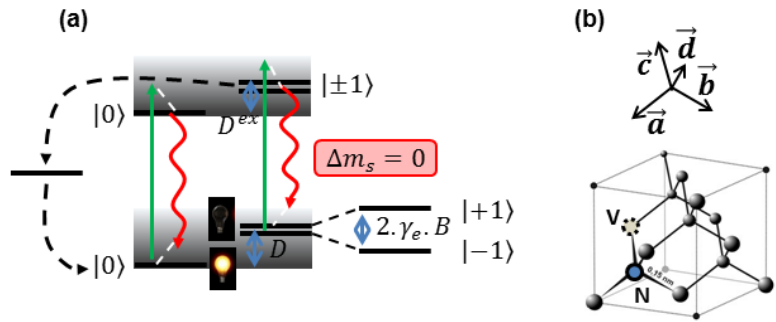

**Figure 5** NV optical properties (b) Simplified energy level diagram of NV-. Ground and excited states are spin triplets. In absence of external magnetic field, the doublet state $|\pm 1\rangle$ is separated from the $|0\rangle$ state due to spin-spin interaction of a value $D = 2.87$ GHz and $D^{ex} = 1.43$ GHz for ground state and excited state respectively [14,15]. Applying a magnetic field

to the NV centre lifts the degeneracy between $|+1\rangle$ and $|-1\rangle$ states due to Zeeman effect of a value $2\gamma_e B$ where $\gamma_e = 2.80$ MHz.G$^{-1}$ is the gyromagnetic ratio of the NV electron spin and $B$ is the projection of the applied magnetic field along NV axis. Solid lines denote optical transitions obeying to the selection rule $\Delta m_s = 0$, the green line corresponding to the green non-resonant excitation and the red to the radiative decay of the NV centre. Dashed lines refer to non-radiative transitions. (b) Representation of the diamond lattice and the four possible orientations named a, b, c and d for NV centre and corresponding to the [111], [1$\bar{1}\bar{1}$], [$\bar{1}$1$\bar{1}$] and [$\bar{1}\bar{1}$1] crystal axis. An example of NV centre oriented along the c direction is given.

In all those experiments, the CVD diamond studied is single crystal which limits to four the possible orientations of a NV centre inside the diamond crystal. More precisely, NV centres can only be oriented along the [111], [1$\bar{1}\bar{1}$], [$\bar{1}$1$\bar{1}$] and [$\bar{1}\bar{1}$1] directions as described on Fig. 5b. In this case, a single ESR measurement give us access to four magnetic field projections values, which can be used to reconstruct the corresponding vector magnetic field [18]. However, unlike ESR measurements on single NV centre where the PL contrast between resonance and off-resonance MW signal is above 20%, for ensembles the contrast is reduced at least four times assuming each direction experiences a different value of magnetic field projection and so possesses different resonance frequencies. Observing a clear signature of ESR with NV ensembles required a setup drastically more stable than for single NV experiments.

To perform ESR measurements through the fibre, a MW antenna was added to the output port of the fibre (Fig. 6a). While sweeping the MW frequency, the PL signal was recorded simultaneously. Results are shown on Fig. 6b. One can observe the clear signature of the eight peaks corresponding to the four possible orientations for the NV centres inside the CVD diamond. Taking advantage of the naturally better coherence properties of NV inside CVD diamond compared to those inside microcrystals, the intrinsic ESR linewidth is small enough to observe the hyperfine structure due to the nuclear spin of Nitrogen atom (I-1) coupled to the NV electron spin. With our system, we managed to achieve a 550 kHz linewidth in CW-ESR measurements decreasing the microwave power to 5 dBm to avoid power broadening which occurred during CW experiments [19]. The magnetic sensitivity can thus be improved using an optically interfaced CVD diamond compared to the case where NV are included into a microcrystal.

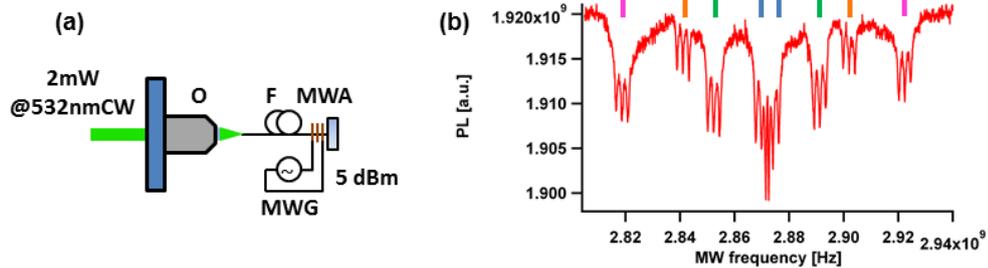

**Figure 6** ESR measurements from a CVD diamond terminated fibre (a) Sketch of the experiment. (b) ESR measurements with 2 mW pump intensity and 5 dBm MW power. Purple, orange, green and blue lines above the graph correspond to peak associated to the same orientation. A magnet was set close to the diamond to split the ESR resonance frequencies.

**4. Summary** To summarize, we have demonstrated a compact and versatile device for magnetic-field imaging where NV centres contained into a CVD diamond are coupled directly to a photonic crystal fibre with a very good coupling efficiency and without using any coupling optics. Taking advantage of the good coherence properties of native NV defects into CVD diamond, we managed to resolve the hyperfine structure linked to interaction between NV electron spin and $^{14}$N nuclear spin. This very simple and versatile system can easily be combined with specifically engineered CVD diamonds when required or even adapted to the investigation of optical properties of other material than diamond.

**Acknowledgements** The authors acknowledge T. Ferhat from NKT Photonics for providing optical fibre required for these experiments, D. Civiale for mechanical parts processing and J.-F. Roch and Michael Trupke for fruitful discussions. This work has been partially funded by the European Community's Seventh Framework Programme (FP7/2007-2013) under the project DIADEMS (grant agreement n°611143).

**References**
[1] J. R. Maze, P. L. Stanwix, J. S. Hodges, S. Hong, J. M. Taylor, P. Cappellaro, L. Jiang, M. V. Gurudev Dutt, E. Togan, A. S. Zibrov, A. Yacoby, R. L. Walsworth, and M. D. Lukin, Nature **455**, 644 (2008).
[2] F. Dolde, H. Fedder, M. W. Doherty, T. Nöbauer, F. Rempp, G. Balasubramanian, T. Wolf, F. Reinhard, L. C. L. Hollenberg, F. Jelezko, and J. Wrachtrup, Nature **7**, 459 (2011).

[3] M. W. Doherty, V. V. Struzhkin, D. A. Simpson, L. P. McGuinness, Y. Meng, A. Stacey, T. J. Karle, R. J. Hemley, N. B. Manson, L. C. L. Hollenberg, and S. Prawer, Phys. Rev. Lett. **112**, 047601 (2014).
[4] D. M. Toylia, C. F. de las Casasa, D. J. Christlea,V. V. Dobrovitskib, and D. D. Awschalom, Proc. Natl. Acad. Sci. USA **110**, 8417 (2013).
[5] I. V. Fedotov, L. V. Doronina-Amitonova, D. A. Sidorov-Biryukov, N. A. Safronov, S. Blakley, A. O. Levchenko, S. A. Zibrov, A. B. Fedotov, S. Ya. Kilin, M. O. Scully, V. L. Velichansky and A. M. Zheltikov, Optics Lett. **39**, 6954 (2014).
[6] I. V. Fedotov, S. Blakley, E. E. Serebryannikov, N. A. Safronov, V. L. Velichansky, M. O. Scully and A. M. Zheltikov, App. Phys. Lett. **105**, 261109 (2014).
[7] I. V. Fedotov, L. V. Doronina-Amitonova, A. A. Voronin, A. O. Levchenko, S. A. Zibrov, D. A. Sidorov-Biryukov, A. B. Fedotov, V. L. Velichansky, and A. M. Zheltikov, Sci. Rep. **4**, 5362 (2014).
[6] G. Q. Liu, X. Y. Pan, Z. F. Jiang, N. Zhao, and R. B. Liu, Sci. Rep. **2**, 432 (2012).
[7] J. R. Maze, A. Dréau, V. Waselowski, H. Duarte, J.-F. Roch and V. Jacques, New Journal of Physics **14**, (2012).
[8] M. Lesik, J.-P. Tetienne, A. Tallaire, J. Achard, V. Mille, A. Gicquel, J.-F. Roch and V. Jacques, Appl. Phys. Lett. **104**, 113107 (2014).
[9] M. Lesik, T. Plays, A. Tallaire, J. Achard, O Brinza, L. William, M. Chipaux, L. Toraille, T. Debuisschert, A. Gicquel, J.-F. Roch and V. Jacques, Diamond and Related Materials **56**, 47 (2015).
[10] T. Ishikawa, K.-M. C. Fu, C. Santori, V. M. Acosta, R. G. Beausoleil, H. Watanabe, S. Shikata and K. M. Itoh, Nano Lett. **12**, 2083 (2012).
[11] M. Fujiwara, H.-Q. Zhao, T. Noda, K. Ikeda, H. Sumiya, S. Takeuchi, Opt. Lett. **40**, 5702 (2015).
[12] N. Alfasi, S. Masis, O. Shtempeluk, V. Kochetok, E. Buks, arXiv preprint arXiv:1601.07718.
[13] L. Childress, M. V. Gurudev Dutt, J. M. Taylor, A. S. Zibrov, F. Jelezko, J. Wrachtrup, P. R. Hemmer, M. D. Lukin, Science **314**, 281 (2006).
[14] G. D. Fuchs, V. V. Dobrovitski, R. Hanson, A. Batra, C. D. Weis, T. Schenkel, and D.D. Awschalom, Phys. Rev. Lett. **101**, 117601 (2008).
[15] P. Neumann, R. Kolesov, V. Jacques, J. Beck, J. Tisler, A. Batalov, L. Rogers, N. B. Manson, G. Balasubramanian, F. Jelezko and J. Wrachtrup, New Journal of Physics **11**, 013017 (2009).
[16] N. B. Manson, J. P. Harrison and M. J. Sellars, Phys. Rev. B **74**, 104303 (2006).
[17] A. Grüber, A. Drabenstedt, C. Tietz, L. Fleury, J. Wrachtrup and C. von Borczyskowski, Science **276**, 2012-2014 (1997).
[18] M. Chipaux, A. Tallaire, J. Achard, S. Pezzagna, J. Meijer, V. Jacques, J.-F. Roch and T. Debuisschert, Eur. Phys. J. D **69**, 7 (2015) 166.
[19] A. Dréau, M. Lesik, L. Rondin, P. Spinicelli, O. Arcizet, J.-F. Roch and V. Jacques, Phys. Rev. B **84**, 195204 (2011).